\documentclass[fleqn,10pt]{wlscirep}
\usepackage{siunitx}             % Added by  RR - For writing units in a consistent style.
\usepackage{graphicx}            % Added by RR
\usepackage{xcolor}              % Added by RR
\usepackage{subfigure}

\begin{document}
%
%\hypersetup{pdfauthor={Ritayan Roy, Paul C. Condylis, Vindhiya Prakash, Daniel Sahagun and Bj\"orn Hessmo}, pdftitle={A minimalistic and optimized conveyor belt for neutral atoms}, pdfkeywords={magneto-optical trap, laser cooling, cold atoms, magnetic trap, magnetic transport, atomic conveyor belt }}
%

%% declare Variance and Mean
%\newcommand{\var}[1]{\mbox{$\mathrm{Var}(#1)$}}
%\newcommand{\mean}[1]{\mbox{$\left\langle #1 \right\rangle$}}
%% declare correlation functions
%\newcommand{\gtwo}[1]{\mbox{$\mathrm{g}^{(2)}(#1)$}}
%% declare a microsecond
%\newcommand{\musec}{\mbox{$\mu{}$s}}
%% declare a micrometer
%\newcommand{\mum}{\mbox{$\mu{}$m}}
%% declare a microKelvin
%\newcommand{\muK}{\mbox{$\mu{}$K}}
%% declare a ket |n>
%\newcommand{\ket}[1]{\mbox{$|#1\rangle$}}

%$^\dagger$

\title{A minimalistic and optimized conveyor belt for neutral atoms}

%\author{Ritayan Roy$^{1, 2, ^\star}$, Paul C. Condylis$^{1}$, Vindhiya Prakash$^{1,3}$, Daniel Sahagun$^{1,4}$, and Bj\"orn Hessmo$^{1, 5}$}
\author[1, 2,*]{Ritayan Roy}
\author[1]{Paul C. Condylis}
\author[1, 3]{Vindhiya Prakash}
\author[1, 4]{Daniel Sahagun}
\author[1, 5]{Bj\"orn Hessmo}

\affil[1]{Centre for Quantum Technologies (CQT), 3 Science Drive 2, Singapore 117543}
\affil[2]{Present address: School of Physics and Astronomy, University of Southampton, Highfield, Southampton, SO17 1BJ, United Kingdom}
\affil[3]{Present address: ICFO- Institut de Ciencies Fotoniques, The Barcelona Institute of Science and Technology, 08860 Castelldefels (Barcelona), Spain}
\affil[4]{Present address: Instituto de F\'{i}sica, Universidad Nacional Aut\'{o}noma de M\'{e}xico, Circuito de la Investigaci\'{o}n Cient\'{i}fica, Ciudad Universitaria 04510 Cd. Mx., Mexico }
\affil[5]{Department of Physics, National University of Singapore, 2 Science Drive 3, Singapore 117542}

\affil[*]{ritayan.roy@u.nus.edu}

%
%% Uncomment for PACS numbers
%\pacs{37.10.De, 37.10.Gh}
%%
%% Uncomment for keywords
%\vspace{2pc}
%\noindent{\it Keywords}: magneto-optical trap, laser cooling, cold atoms, magnetic trap, magnetic transport, atomic conveyor belt
%

\begin{abstract}

Here we report of a design and the performance of an optimized micro-fabricated conveyor belt for precise and adiabatic transportation of cold atoms. A theoretical model is presented to determine optimal currents in conductors used for the transportation. We experimentally demonstrate a fast adiabatic transportation of Rubidium\,($^{87}$Rb) cold atoms with minimal loss and heating with as few as three conveyor belt conductors. 
This novel design of a multilayered conveyor belt structure is fabricated in aluminium nitride\,($AlN$) because of its outstanding thermal and electrical properties. This demonstration would pave a way for a compact and portable quantum device required for quantum information processing and sensors, where precise positioning of cold atoms is desirable.

%We report of a design and the performance of an optimised micro-fabricated conveyor belt for precise and adiabatic transportation of ultra cold atoms. A theoretical model is presented to determine the least number of conveyor conductors required as well as for the optimisation of current through them for an adiabatic transportation. We experimentally demonstrate with as few as three conveyor conductors a fast adiabatic transportation of Rubidium\,($^{87}$Rb) cold atoms with minimal loss. The minimized number of conveyor conductors ensures a reduction in the number of power supplies required for the transportation of cold atoms as well as simplifies the fabrication process due to fewer conductors. It also reduces the complex electrical connections inside the vacuum chamber. This novel design of a multilayered conveyor belt structure is fabricated in aluminium nitride\, ($AlN$) due to its outstanding properties such as high thermal conductivity, low electrical resistivity and high hardness. This demonstration would pave a way for a compact and portable quantum device required for quantum information processing and sensors, where precise positioning of the cold atoms is indispensable. 

\end{abstract}

\flushbottom
\maketitle
\thispagestyle{empty}

\section*{Introduction}

A simple yet powerful technique to reach the quantum regime with neutral atoms has been demonstrated using atom chips \cite{ReichelMirrorMOT, Fol00}. The micro-fabricated wires and electrodes are lithographically imprinted on the atom chip to construct complex potentials using magnetic and electric fields to trap and manipulate the atoms \cite{Folman:ChipReview, reichel2011, Reichel:2002, Fortagh2007}. Many aspects of integrated matter wave technology have been demonstrated such as combined magnetic/ electrostatic traps \cite{Kru03}, transportation of neutral atoms \cite{Han01}, atomic beam splitters \cite{Cas00}, the creation of Bose-Einstein condensation (BEC) \cite{Ott01,Hae01,Sch03}, atom interferometer \cite{Sch05, Bohi:2009ku}, and the integration of optical \cite{Gallego:2009ww} and magnetic \cite{leung2011, jose2014} lattices with atom chip.

For many applications, such as in quantum information and communication, quantum sensors and microscopes, and to study various short and long range interaction potentials, it is desired to precisely move trapped atoms  in a three-dimensional (3-D) geometry along with temporal control. Indeed the realisation of a ``qubit conveyor belt" is required for many implementations of quantum computing \cite{DiVincenzo:2000dn}.

In this regard the transportation of ultra cold atoms using a magnetic ``conveyor belt" should be adiabatic such that the atoms remain in the same state during the transport. The demonstrations of transportation of neutral atoms using a magnetic conveyor belt has been realised on an atom chip with two different patterns of conveyor belt wires. The first, using two wires in a repeating square wave type pattern on the surface of the atom chip \cite{Han01}. It is demonstrated that, by applying time dependent currents to the wires, the atom cloud could be split and transported parallel to the atom chip surface. In the second, an array of eight wires are used repeating on the chip surface. By manipulating the currents in those eight wires the atom cloud is transported \cite{Gunther2005, Gierling:2011ta}. Both these implementations have some advantages and drawbacks. In the first, using only two wires has the distinct advantage that wires can be placed in one plane. However, it won't be easy to translate the trap without imperfections by modifying only the currents. Thus, adiabatic transport is difficult to achieve. In the second experiment, the use of a set of eight repeating wires would allow to eliminate even higher order trap imperfections by exploiting our approach. The implementation using eight wires however requires twice the number of power supplies compared to the scheme presented here.
%Furthermore connecting the wires through to power supplies outside the vacuum chamber would be cumbersome. 

%In the first, the transportation of atoms was not completely adiabatic. In the second, the use of a set of eight repeating wires presents some limitations in terms of chip fabrication, and connecting the wires through to power supplies outside the vacuum chamber is cumbersome. We have demonstrated a order of magnitude faster adiabatic atom transport (40 mm/s) in comparison with the previous works which were 5.3 mm/s \cite{Han01} and 2.6 mm/s \cite{Gunther2005, Gierling:2011ta}.

In this article a theoretical model is presented to determine both the least number of conductors required for a conveyor belt and the optimised current through them to ensure adiabatic transportation of atoms. We present a new atom chip based conveyor belt design, which is simple to fabricate using printed circuit board\,(PCB) fabrication techniques. To achieve the adiabatic performance, our design utilises a repeating pattern of just four wires, though only three wires are strictly required. We outline the basic design of our conveyor belt and we experimentally demonstrate the adiabatic performance of the conveyor belt. With as few as three conveyor conductors, adiabatic transport of atoms with minimal loss at a maximum velocity of $v_{max}$= 40 mm/s is demonstrated. This is almost an order of magnitude faster transport in comparison with the previous works as reported to be 5.3 mm/s \cite{Han01} and 2.6 mm/s \cite{Gunther2005, Gierling:2011ta}.  

%We have found the heating rate in terms of heating per millimetre of transport to be $36\pm24$ nK.mm\textsuperscript{-1}. 

\section*{Methods}
\label{sec:design}

%\begin{figure}[t]
%\centering
%\subfigure[] {\includegraphics[width=0.4\textwidth]{1.pdf}\label{AtomChipPad}}
%%\subfigure[] {\includegraphics[width=0.45\textwidth]{CB-BaseChipZoom.pdf}\label{CB-BaseChipZoom}}
%\subfigure[]{\includegraphics[width=0.7\textwidth]{2.pdf}\label{CB-BaseChip}}
%\subfigure[]{\includegraphics[width=0.45\textwidth]{3.pdf}\label{ChipWithHolder}}
%
%\caption{The Science and Base chip schematic. The science chip, a), is comprised of an array of wires to produce different magnetic trap geometries. The central wire, which runs the full length of the chip produces the guide potential for the magnetic conveyor belt. We refer to this wire as the ``I wire", given its shape. Mounted below the science chip is a base chip comprising multiple layers. The second layer, b) has an array of wires which we use to produce the magnetic conveyor belt to move the atoms along the science chip surface. There are 4 repetitive wires, seperated by 400 $\mu m$ from each-other. The AIN base chip, c), has 14 layers. The surface layer has pads to connect the science chip wires, right, and pads for copper wire connections to a feedthrough, left. Copper vias run down through the chip and allow wires to be connected on multiple layers. The conveyor belt wires are connected to each other using the layer below. The fourth layer connects the science chip wires to the surface connection pads. To allow higher currents to flow we repeat this structure on several layers \cite{Roy16}.}
%  \label{fig:conveyor} 
%\end{figure}

\begin{figure}[tb]
\centering
\subfigure[] {\includegraphics[width=0.45\textwidth]{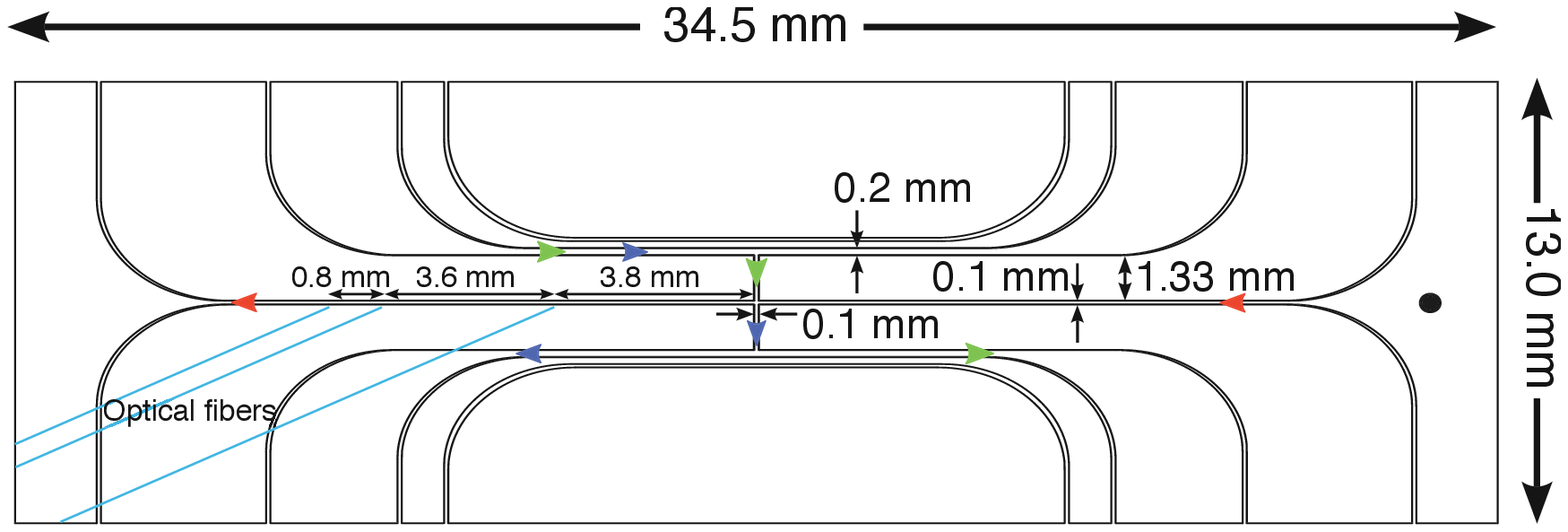}\label{AtomChipPad}}
%\subfigure[] {\includegraphics[width=0.45\textwidth]{CB-BaseChipZoom.pdf}\label{CB-BaseChipZoom}}
\subfigure[]{\includegraphics[width=0.65\textwidth]{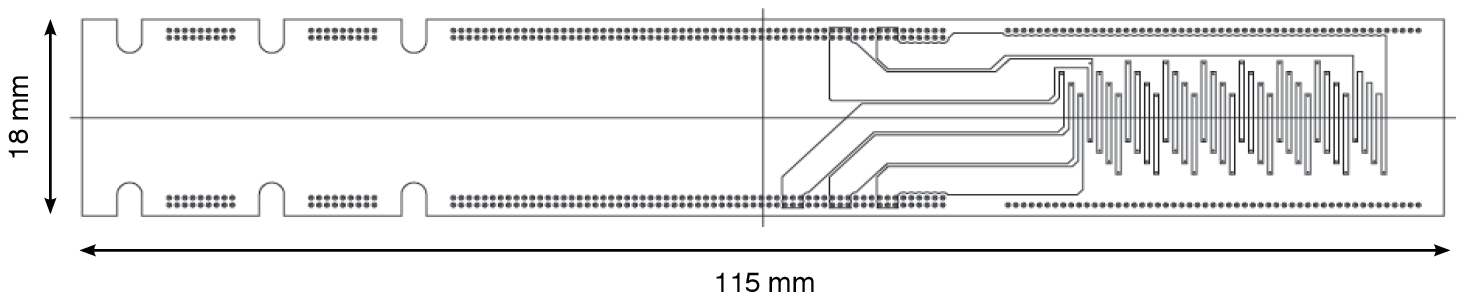}\label{CB-BaseChip}}
\subfigure[]{\includegraphics[width=0.40\textwidth]{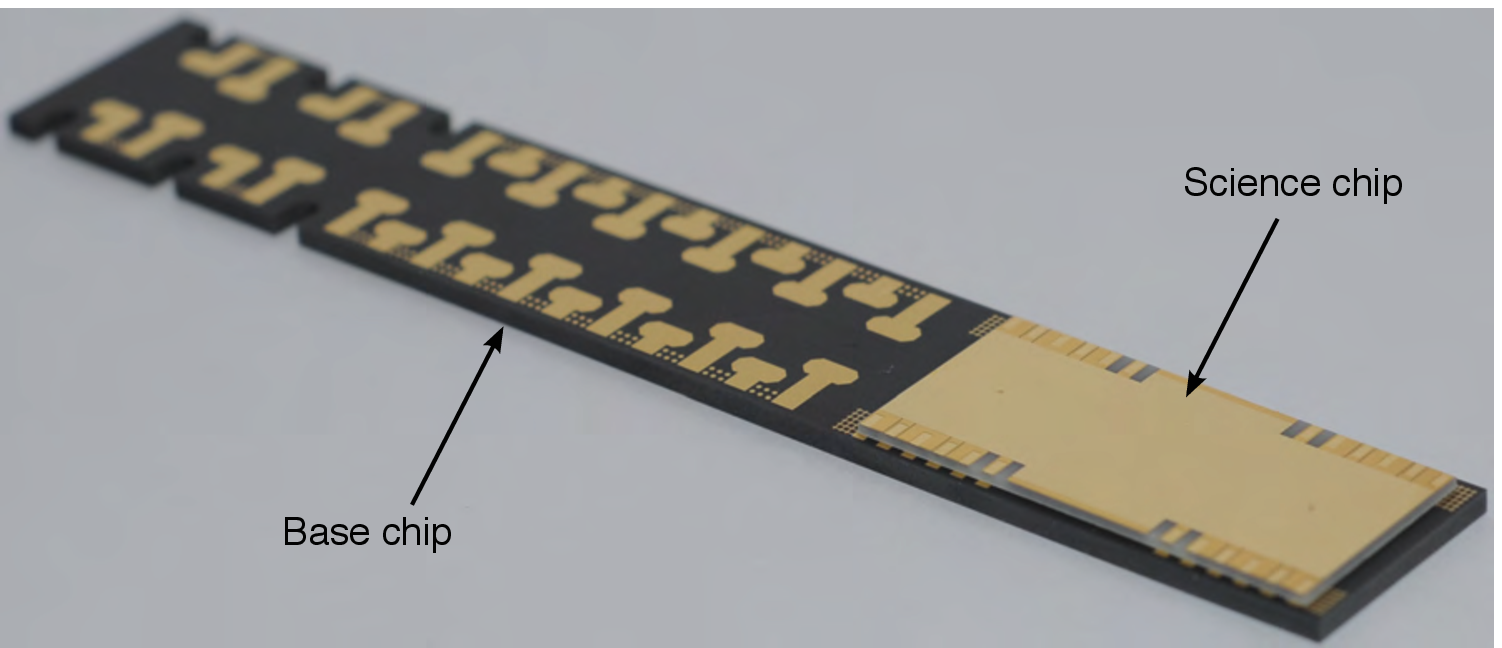}\label{ChipWithHolder}}

\caption{The science and base chip schematic. The science chip, (a), is comprised of an array of wires to produce different magnetic trap geometries. The wires are named after characters in the alphabet they geometrically resemble. The direction of current flow through each wires is indicated with arrows. In combination with external magnetic fields, the current through the ``U-wire"\,(indicated with blue arrows) generates the quadrupole field for the magneto-optical trap, the ``Z-wire"\,(indicated with green arrows) is used for the magnetic trap, and the ``I wire"\,(indicated with red arrows) produces the magnetic guide potential for the magnetic conveyor belt. The thickness of the science chip is around 530 $\mu$m. Mounted below the science chip is the base chip. The base chip is made of $AlN$ and has 13 layers. The second layer, (b), has an array of wires which are used to produce the magnetic conveyor belt. There are four repeated wires, separated by 400 $\mu$m from each-other. The surface layer, (c), has pads to connect the science chip wires\,(right), and pads for copper wire connections to a feedthrough\,(left). The pads are connected with internal conductors. To allow higher currents we repeat this structure over several layers. The conveyor belt wires are located around 120 $\mu$m below the top surface of the base chip. The total thickness of the base chip is 1.5 mm.}

  \label{fig:conveyor} 
\end{figure}

Our atom chip is comprised of a two component system with a science chip, connected underneath to a base chip. On the science chip, lithographically printed gold micro-wires in the shape of ``U, Z, and I" are used\,(Figure \ref{fig:conveyor}(a)) to create different magnetic field geometries to trap and guide the atom cloud. The ``I wire", the central wire running through the entire length of the science chip, is used along with the conveyor wires on the base chip for the transportation of the cold atom cloud. All of the science chip wires are connected to the base chip by conductive epoxy.

 %%%%%%%Moved here from Methods section%%%%%%% %%%%%%%Moved here from Methods section%%%%%%%
 %%%%%%%Moved here from Methods section%%%%%%% %%%%%%%Moved here from Methods section%%%%%%

The base chip performs two functions; first, the base chip has its own set of wires, which form the conveyor belt transport wires as shown in the Figure \ref{fig:conveyor}(b) and second, to connect the science chip to copper wires that exit the vacuum chamber via a vacuum feedthrough as shown in the Figures \ref{fig:conveyor}(c). These two functions are enabled by using a multilayer base chip design, fabricated in aluminium nitride\,($AlN$). $AlN$, a covalently-bonded ceramic, is chosen for fabrication due to its outstanding properties such as high thermal conductivity and high electrical resistivity.  The science chip is designed and manufactured in-house \cite{Roy16}. The base chip is manufactured by commercially available PCB fabrication methods techniques used for the heat sinks of LED and CMOS chips. 
%A full description of the design and manufacturing of the science and base chip can be found in this thesis \cite{Roy16}.

The experimental sequence for the cold atom transport is as follows. The \textsuperscript{87}Rb atoms are magneto-optically trapped\,(MOT) using a bias magnetic field and current through a U-shaped copper wire\,(U-wire) of 3 $\times3$ mm cross-section, approximately 4 mm directly below the U-wire on the science chip\,(and base chip), using the well known mirror-MOT technique \cite{ReichelMirrorMOT, Fol00}. The science chip's gold\,(Au) surface acts as a mirror for the MOT. The current in the copper U-wire\,(not shown), initially 24 A for the MOT loading time for 10 seconds, is ramped down while the science chip's U-wire (Figure \ref{fig:conveyor}(a)) current is ramped up, to a maximum of 1.5 A. In this way approximately 1.5 $\times10^6$ atoms are transferred to the MOT created by the science chip's U-wire. A typical temperature of around 200$\mu$K is observed . The magnetic field gradient is calculated to be around 10-15 G/cm for the MOT loading stage.

The atoms are then compressed to mode-match with the magnetic trap created by the Z-wire\,(Figure \ref{fig:conveyor}(a)) using a magnetic field gradient of 55 G/cm and are held 350 $\mu$m away from the chip surface. A short stage\,(5 ms) of optical molasses cooling is performed after the compression of the MOT, followed by an optical pumping before loading into the magnetic trap created by the Z-wire of the science chip\,(Z-MT). All the magnetic fields are turned off during the optical molasses cooling.

To improve the loading efficiency of the magnetic trap, all the atoms from different m$_{F}$ levels of the MOT are optically pumped to the 5$^{2}S_{1/2}$, F\,=\,2, m$_{F}$\,=\,2 weak-field-seeking state. A circularly polarized $\sigma$\,+ beam is used for optical pumping and a magnetic bias field of 3 G along the beam direction is used to define the quantization axis. After the optical pumping, the Z-wire current is ramped up from 0 to 1.5 A. At the same time the magnetic field perpendicular to the I-wire is ramped up to 1 G, while the magnetic field along the I-wire is kept at 10 G to load the atoms in the Z-MT.

The atoms are transferred from Z-MT to a ``dimple" magnetic trap using the guide wire\,(I-wire) on the science chip and the conveyor belt wires in the base chip. Typically 300 $\times10^3$ atoms are loaded into the trap, though it is possible to capture up to 1 $\times10^6$ atoms from the molasses. The temperature of the atom cloud in the magnetic trap is measured around 10 $\mu K$. The current through the I-wire is 0.9 A, the magnetic fields parallel and perpendicular to the I-wire are 5.4 G and 18.2 G respectively\,(see Figure~\ref{fig:3wires}). These values are kept constant during the atom transportation and only the currents passing through the conveyor wires are altered. The current through one of the conveyor wires, which is below the Z-wire, is kept at 0.5 A for the dimple trap loading. Current through the individual conveyor wire is limited to 1.5 A. After loading the conveyor belt using a dimple trap, we transport the atoms to a set of on-chip fibre optics as indicated in the Figure \ref{fig:conveyor}(a). In order to minimise heating and atom loss from the trap during transport, we implement optimized current waveforms through the conveyor wires. The optimization technique is described in the section below.

\section*{Results}
\subsubsection*{Optimization of the currents}
\label{sec:optimization}

For a smooth transport along the wire it is required that the trapping frequencies remain the same at all positions along the wire. To optimise the conveyor belt, we use a one dimensional model where we assume that the transverse and longitudinal trapping frequencies are independent. This approximation is valid when the principal axes of the trap are aligned with the trapping wires. 

\begin{figure}[tb]
    \centerline{\includegraphics[width=0.65\textwidth]{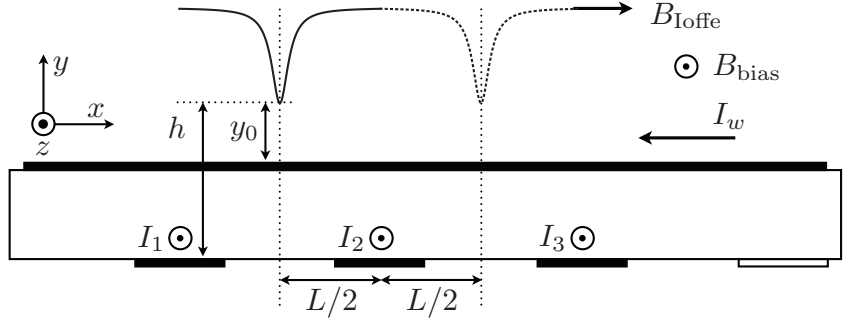}}
    \caption{\label{fig:3wires} The conveyor belt consists of a guide wire carrying a current $I_w$, and transport wires with currents $I_1$, $I_2$, and $I_3$, etc. to the last wire $I_N$. The transport wires are separated a distance $L$ from each other. The magnetic fields from these wires are counteracted by external magnetic fields to $B_{\mathrm{Ioffe}}$ and $B_{\mathrm{bias}}$. }
\end{figure}

Consider the situation illustrated in Figure~\ref{fig:3wires} where atoms are held by a guide wire along the x-axis, carrying a current $I_w$. A guide is formed when an external bias field $B_{\mathrm{bias}}$ is applied perpendicular to the wire. The guide will be located at a height,
\begin{eqnarray}
\label{trapbottom}
 y_0=\frac{\mu_0}{2\pi}\frac{I_w}{B_{\mathrm{bias}}}.
 \end{eqnarray}
In such a guide the potential is a quadrupole field, where atoms would be lost due to the Majorana spin flips. Applying an additional field along the guide wire $B_{\mathrm{Ioffe}}$ will remove this zero-point of the field. For this guide the transverse trapping frequency is given by,
$
\omega_\perp=\sqrt{\mu_B g_F\frac{d^2B}{dr^2}/M}
$
where $M$ is the mass of the atom and $g_F$ the Land\'e factor, $\mu_B$ the Bohr magneton, and $r$ the distance from the minimum. 
The transverse field curvature around the minimum is given by,
$$
\frac{d^2B}{dr^2}=\left(\frac{2\pi}{\mu_0}\right)\frac{B_\mathrm{bias}^2}{I_w\sqrt{B_\mathrm{Ioffe}}}.
$$
At the centre of the guide the magnetic field is equal to $B_{\mathrm{Ioffe}}$.

Trapping along the guide is formed by running currents ($I_k$, $k=1,2,3$) through wires perpendicular to the guide wire that counteract and reduce the field $B_{\mathrm{Ioffe}}$. Assuming that $y_0$ remains unaffected by the wires $I_k$, the field along the guide centre becomes
\begin{eqnarray*}
B_x&=&B_{\mathrm{Ioffe}}
-\frac{\mu_0 I_1}{2\pi}\frac{h}{h^2+(x+L)^2}
-\frac{\mu_0 I_2}{2\pi}\frac{h}{h^2+x^2}
-\frac{\mu_0 I_3}{2\pi}\frac{h}{h^2+(x-L)^2},
\end{eqnarray*}
where, $L$ is the separation between the wires, and $h$ the distance from the wires to the guide as illustrated in Figure~\ref{fig:3wires}. It is assumed that the wires are thin and long.

Here we calculate the currents that allows a smooth transport of atoms from $-L/2$ to $L/2$. 
When the atoms have reached $L/2$, the process can be repeated by shifting the current configurations to new wires. 
During the transport  the following conditions should be satisfied: i) The current configuration provides a single minimum for $B_x$ at any position between $-L/2$ and $+L/2$, ii) the magnetic field minimum should remain constant to maintain a constant transverse trap frequency, and iii) the axial  curvature should remain constant to keep the axial trap frequency constant. These three conditions allow us to solve for the three currents.
The equations are:
\begin{eqnarray}
\label{condition1}
&& \left.\partial_x B_x(x)\right|_{x=x_0} = 0\\
\label{condition2}
&&B_x(x_0)=B_0<B_{\mathrm{Ioffe}} \\
\label{condition3}
&&\left.\partial_{xx} B_x\right|_{x=x_0} =C, 
\end{eqnarray}
where, $B_0$ is the wanted trap bottom, and $C$ is the desired field curvature.

\begin{figure}[tb]
    \centerline{\includegraphics[width=0.8\textwidth]{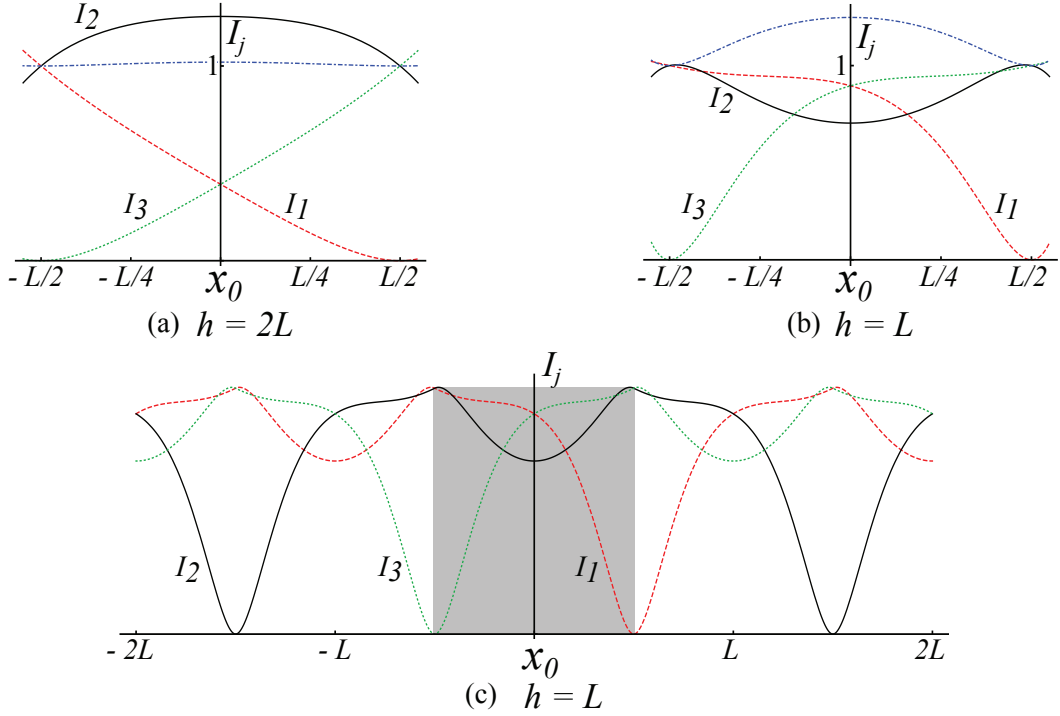}}
    \caption{\label{fig:currents} In (a) and (b) we plot the currents $I_1$\,(red), $I_2$\,(black), and $I_3$\,(green) as the trap position is moved from $-L/2$ to $L/2$. For the position $-L/2$ the currents are normalised to $I_1=I_2=1$. The blue curve is the total current divided by two. In (a) for $h=2L$, the total current in the three wires is almost constant. In (b) for $h=L$, more total current is required to maintain a constant axial trap  frequency. It will not be possible to transport atoms if $h<\sqrt{3/4}L$. In (c) the current waveforms provided by several wires separated by a distance $L$ have been stitched together to provide a longer transport distance. The behaviour of the equations (\ref{waveform1}-\ref{waveform3}) are illustrated without units in this figure.}
\end{figure}

Solving these equations for $I_1$, $I_2$ and $I_3$ gives:
%\begin{widetext} %%%NJP
\begin{eqnarray}
\label{waveform1}
&I_1&=
-\frac{\pi\left(h^2+\left(L+x_0\right)^2\right)^3}{R} \left[C \left(h^2+x_0^2\right) \left(h^2+\left(L-x_0\right)^2\right) \left(h^2+x_0
   \left(L-x_0\right)\right)\right.\nonumber\\
&&   \left.-2\left(B_{\mathrm{Ioffe}}-B_0\right) \left(h^4+h^2 \left(L^2-6 L x_0+6 x_0^2\right)-3 x_0^2 \left(L-x_0\right)^2\right)\right]\\
\label{waveform2}
&I_2&=
\frac{2\pi\left(h^2+x_0^2\right)^3}{R} \left[C \left(h^2+\left(L-x_0\right)^2\right) \left(h^2+L^2-x_0^2\right)
   \left(h^2+\left(L+x_0\right)^2\right)\right.\nonumber\\
&&   \left.-2 \left(B_{\mathrm{Ioffe}}-B_0\right) \left(h^4-2 h^2 \left(L^2-3 x_0^2\right)-3
   \left(L^2-x_0^2\right)^2\right)\right]\\
\label{waveform3}
 &I_3&=
 -\frac{\pi\left(h^2+\left(L-x_0\right)^2\right)^3}{R} \left[C \left(h^2+x_0^2\right) 
   \left(h^2+\left(L+x_0\right)^2\left(h^2-x_0 \left(L+x_0\right)\right)\right)\right.\nonumber\\
&&   \left.-2 \left(B_{\mathrm{Ioffe}}-B_0\right)\left(h^4+h^2 \left(L^2+6 L x_0+6 x_0^2\right)-3 x_0^2
   \left(L+x_0\right)^2\right)\right],
\end{eqnarray}
where, 
$$
R=2\mu_0 h
   L^2 \left[5 h^6+h^4 \left(6 L^2-9 x_0^2\right)+h^2 \left(L^4-12 L^2 x_0^2+15 x_0^4\right)-3 \left(x_0^3-L^2 x_0\right)^2\right]
$$

It is convenient to select a simple starting condition, for instance $I_1=I_2$ and $I_3=0$. For this starting condition the curvature becomes 
$$
C=\frac{8(4h^2-3L^2)}{(4h^2+L^2)^2}\left(B_{\mathrm{Ioffe}}-B_0\right).
$$
This also gives an important constraint on the geometry. For a wire separation $L>\sqrt{4/3}h$ it will not be possible to form a field minima at $\pm L/2$. At the end of the transport sequence the current configuration will be $I_2=I_3$ and $I_1=0$. To repeat the transport across multiple transport wires, this final configuration is the starting condition for transport using wires $I_2$, $I_3$, and $I_4$ located at a distance $L$ from wire $I_3$. 
The currents for this is plotted in Figure~\ref{fig:currents}.

\begin{figure}[tb]
\centering{\includegraphics[width=0.8\textwidth]{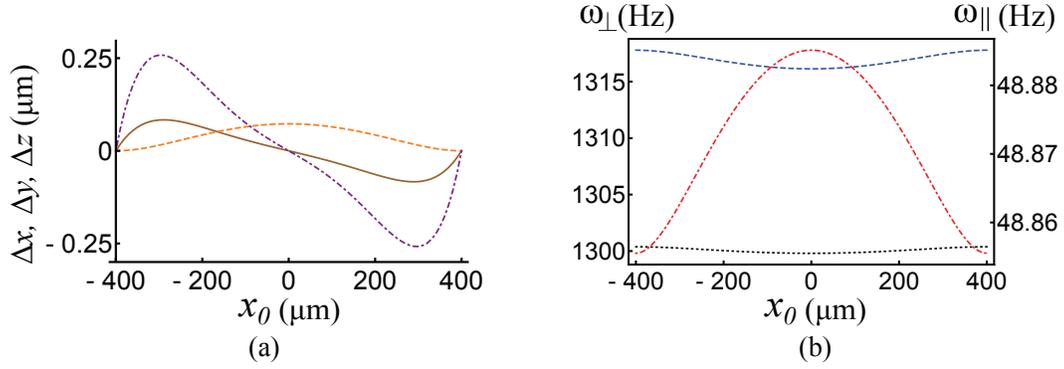}}
 \caption{In (a) we plot the numerically calculated deviation in position of the atoms along the three axes during transportation along the x axis. We scan $x_0$ , the trap centre, from $-L/2$= -400 $\mu m$ to $L/2$= 400 $\mu m$. The brown line shows $\Delta x=x_\mathrm{exact}-x_0$. This deviates by less than 100 nm from the wanted value. The orange line shows  $\Delta y=y_\mathrm{exact}-y_0$, where, $y_0$ is the vertical distance to the guide wire for $x_0=-400 \mu m$. This height changes about 100 nm. The purple line shows $\Delta z$, the deviation along the $z$ axis. This deviation is supposed to be zero if the approximation is exact. Here we note that the trap wiggles about 250 nm along the chip surface which is insignificant for a transport distance of 800 $\mu m$. In (b) the transverse\,(black and blue) and axial trap\,(red) frequencies are plotted as the trap is moved from -400 $\mu m$ to 400 $\mu m$. From the numerical calculation, we observe that the trap frequencies are almost constant during atom transport. The axes representation follows from the Figure~\ref{fig:3wires}. }
\label{fig:numeric test} 
\end{figure}

At distances $h\gg L$ these expressions simplify further:
\begin{eqnarray*}
I_1&=&\frac{I_0}{2}\left(\frac{1}{2}-\frac{x_0}{L}\right)^2\\
I_2&=&\frac{I_0}{2}\left(\frac{3}{2}-\frac{2 x_0^2}{L^2}\right)\\
I_3&=&\frac{I_0}{2}\left(\frac{1}{2}+\frac{x_0}{L}\right)^2,
\end{eqnarray*}
where, $I_0=I_1+I_2+I_3$ is the total current through the wires. For the above calculations, we used an approximation that assumes the the conveyor belt wires do not influence the transverse position of the trap. 

To test this approximation, we calculate the magnetic fields by integrating Biot-Savart's law along the current carrying wires. Here we use the parameters close to our experimental values, i.e., $L$= 800 $\mu$m, $I_w$= 0.9 A, $I_k$= 0-1.5 A, bias field= 5 G and Ioffe field= 10 G. The basic requirement for this approximation to work is that $h\gg y_0$. From the numerical test, we have found that the trap wiggles about 250 nm along the chip surface which is insignificant for a transport distance of 800 $\mu m$ as shown in the Figure~\ref{fig:numeric test} (a) and both axial and radial trap frequencies are almost constant during atom transport as shown in the Figure~\ref{fig:numeric test} (b).

%%

%
%\begin{figure}[tb]
%    \centerline{\includegraphics[width=0.5\textwidth]{test3.pdf}}
%    \caption{\label{fig:numeric test} In a) we plot the numerically calculated position as we scan $x_0$ from $-L/2=-400\mu m$ to $L/2=400\mu m$.
%    The solid line shows $\Delta x=x_\mathrm{exact}-x_0$. This deviates by less than 100nm from the wanted value. The dashed line shows  $\Delta y=y_\mathrm{exact}-y_0$, where $y_0$ is the distance to the guide wire for $x_0=-400\mu m$. This height changes about 100nm. The dotted line shows $\Delta z$. This number is supposed to be zero if the approximation is exact. Here we note that the trap wiggles about 250nm along the chip surface. In b) the transverse (dashed lines) and axial trap (solid line) frequencies are plotted as the trap is moved from from $-L/2=$ to $L/2$.
%    }
%\end{figure}
%%

%%

\subsection*{Conveyor belt characterisation}

\label{sec:characterisation}

To verify the model described in the above section, we have performed temperature measurements of the magnetically trapped atoms after transportation. An important parameter of the current waveforms is the height ($h$), which is the distance between the atom cloud and the conveyor belt wires. It is evident from the Figure \ref{fig:currents} that the waveforms change significantly depending on this height ($h$), as provided in the equations (\ref{waveform1}-\ref{waveform3}). The magnetic trap minimum in the vertical direction perpendicular to the chip surface is determined  by the guide wire current and the external bias field as provided in equation \ref{trapbottom}. 

In order to move the atoms with a minimal change in curvature, the height ($h$) should be set correctly in equations (\ref{waveform1}-\ref{waveform3}). 
In reality the exact value for the height of the cloud from the conveyor belt is not known. We experimentally determine true the value of $h$ in the following manner. We transport the atoms using current waveforms calculated for different choices of $h$. If the real distance between the cloud and the surface matches one of our choices of $h$, we would expect to see minimal heating of the cloud during transportation as that would satisfy all the three conditions given in equations \ref{condition1}, \ref{condition2}, and \ref{condition3}. However, if the value of $h$ is incorrectly assigned in the calculation, we would expect heating. 

%In the experimental setup the height of the cloud from the conveyor wire is not exactly known, so, we guess the value of $h$ from the calculated value of $y_{0}$ using equation \ref{trapbottom} and the thickness of the science chip. If the guess value of the cloud to the conveyor wire distance ($h$) is matched, then the current through the conveyor wires would be optimized. Optimized current through the conveyor wire would lead to a minimum heating of the cloud during transport as that would satisfy all the equations \ref{condition1}, \ref{condition2}, and \ref{condition3} . However, if the value of $h$ is incorrectly assigned in the calculation we would expect heating of the cloud due to changes of the curvature of the trap. 

To verify this we assign several values for $h$ from 0.8 mm to 4.0 mm and calculate different current waveforms for the conveyor wires. For each guessed value of $h$, we measure the temperature of the cloud after the transportation. In this experiment the atoms are moved a distance of $1.6$ mm\,($0.8$ mm in the forward direction and again backwards). 
To eliminate any heating due to changing the loading conditions of the cloud, we perform a differential temperature measurement. We measure the difference in temperature of the cloud that is transported versus a stationary cloud loaded into the magnetic-trap under the same conditions. This differential measurement thus reveals the heating solely due to the transport. 
%Changing the height used in the waveforms changes the initial current in the conveyor wires used to load the atoms from the molasses. 

\begin{figure}[tb]
\centering
    \includegraphics [width=0.8\textwidth]{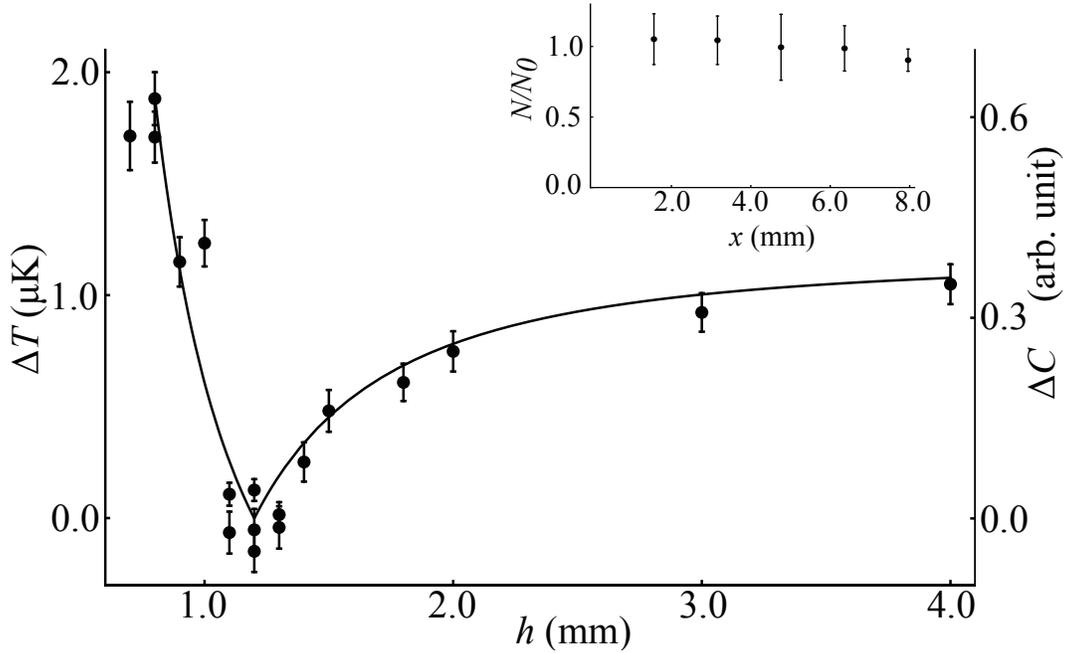}
    \caption{\label{fig:ConveyorTemp} Heating due to the transportation of the cloud as a function of the current waveform height parameter. The minimum heating occurs at $h$=1.2 mm. The solid line shows the total change in curvature ($\Delta C$) of the magnetic trap due to transport, calculated numerically as a function of $h$. We clearly observe a similar behaviour of the cloud heating and the curvature change. This validates that the heating is due to changes in the curvature of the trap at different location along the transport route. Inset: Atom loss due to transportation. $N/N_0$ is the relative atom number remaining after transport, compared to a stationary cloud held for the same time. The points are an average of three measurements, and the error bar is the standard deviation. A little loss is observed after 6.4 mm of transport, and the loss is most likely due to collision of atoms with optical fiber placed on the chip surface as shown in the Figure \ref{fig:conveyor}(a). The atoms are transported from the center of the science chip towards the fiber.}
\end{figure}

Figure \ref{fig:ConveyorTemp} shows that there is almost no heating of the cloud for $h$= 1.2 mm. This is consistent with the distance between the cloud and the conveyor belt wires from the theoretical prediction. The distance of the cloud from the science chip surface ($y_0$) is estimated to be around 100 $\mu m$\,(equation \ref{trapbottom}), where, the current through the guide wire is 0.9 A along with a magnetic field of 18.2 G perpendicular to the wire. Each data point corresponds to the mean difference of three measurements of the temperature for the moved and stationary clouds. The error bars indicate the standard deviation. For the values of $h$= 1.1, 1.2, and 1.3 mm, additional measurements are performed to verify the repeatability. 

%In the Figure, the solid line shows the total change in curvature of the magnetic trap along the transport route. This is calculated numerically from the 3-dimensional trap geometry given the conveyor belt currents, $I_1, I_2, I_3$, the guide-wire current, $I_w$, and the bias fields, $B_{\mathrm{bias}}$ and $B_{\mathrm{Ioffe}}$ respectively. 

We observe that the heating follows the changes in curvature of the trap as shown in the Figure \ref{fig:ConveyorTemp}, as expected. For the correctly assigned value of $h$, no measurable heating is observed. The trapping frequencies remain almost constant at all positions as predicted in the Figure~\ref{fig:numeric test} (b). At large cloud-wire distances the current waveforms change very little, however at distances very close to the wires the waveforms change dramatically, as shown in the Figure \ref{fig:currents}. This explains the plot of the differential temperature measurement in Figure \ref{fig:ConveyorTemp}, where we observe greater heating of atoms closer to the conveyor wire than further away. We didn't observe any increase in temperature within our statistical limits at $h$= 1.2 mm, during the transportation.  

Another important performance criterion for an atom chip conveyor-belt, is to demonstrate that atoms are not lost due to the transport. When atoms are trapped in a magnetic potential, atoms are lost over time, due to collisions of background gas in the vacuum system. In order to confirm if the atom loss occurs during the transportation using the the conveyor-belt, one needs to take into account the atom loss due to the magnetic trap's lifetime. To verify, we measure the atom number for a transported cloud and for a stationary cloud held for same time as the transportation time, with the exactly same loading conditions. We then divide the measured atom numbers from each other and observe the atom loss due to the transport, independent of the natural trap lifetime. The measurement is then repeated over different distances of transport. The inset in Figure \ref{fig:ConveyorTemp} shows the relative atom number over different transport distances. The points in the plot are an average of three measurements, with the standard deviation forming the error bars. We observe no atom loss for distances up to 6.4 mm, and 91\% of the atoms remain at 8 mm. This small loss is most likely due to some collisions with an optical fiber placed on the chip surface. 

For an ideal conveyor belt one would like to move the atomic cloud not only adiabatically, but also as quickly as possible in order to minimise dead time and to avoid atom loss due to magnetic trap's limited lifetime. Once the atom cloud to the conveyor wire distance\,($h$) is found, it is possible to move the atoms with any position-acceleration profile, limited only by the gradient and depth of the trap. We have found no significant heating dependent on the acceleration profile. In this experiment, the atoms are transported adiabatically almost without loss, with a maximum velocity of $v_{max}$= 40 mm/s.

\section*{Discussion and Conclusion}
\label{conclusion}

We have shown an analytical model of a minimalist magnetic conveyor belt for neutral atoms trapped using atom chip technology. Using this model we have calculated the optimal set of wire currents needed to transport atoms across the surface without changing the trap curvature or height, and thus maintaining the temperature of the atom cloud unchanged. The theoretical model and current optimization technique for adiabatic atom transport is not limited to a magnetic trap, this model could also be adapted for ion transport. 
We have built and characterised the conveyor belt, showing almost no heating or loss due to the transportation of the trapped atoms at a maximum velocity of $v_{max}$= 40 mm/s.  Combining our method with more wires \cite{Gunther2005, Gierling:2011ta}, we expect it to be possible to perform ultra-quiet transport, or to accelerate atoms to very high velocities. The fast transport allows to overcome the atom loss due to the life time, as well as perform faster measurements. The minimalistic and optimized approach, towards the design of the conveyor belt, lends itself for easy implementation towards compact and portable atom chip devices required for quantum sensors, quantum optics and information.

%---------------------

\section*{Acknowledgements}
The authors would like to thank A. Dhanapaul, S. M. Maniam, and J. Andersson for fabricating the science chip, and for many valuable and insightful discussions. We also thank J. Yik, R. Srinivas, and N. Kia Boon for their work on the experiment. This research has been supported by the National Research Foundation \& Ministry of Education, Singapore. 

\section*{Author contributions statement}
R.R. and P.C. made the experimental setup, took the data, analysed the data. V.P. and D.S. designed the base chip and performed testing along with R.R. and P.C.  B.H. did the theoretical modelling and supervised all the work. All authors contributed to the discussions and given valuable inputs in the manuscript preparation. All authors reviewed the manuscript.

\section*{Additional information}
\textbf{Competing financial interests: } The authors declare that they have no competing interests.


\begin{thebibliography}{10}
\expandafter\ifx\csname url\endcsname\relax
  \def\url#1{\texttt{#1}}\fi
\expandafter\ifx\csname urlprefix\endcsname\relax\def\urlprefix{URL }\fi
\expandafter\ifx\csname doiprefix\endcsname\relax\def\doiprefix{DOI }\fi
\providecommand{\bibinfo}[2]{#2}
\providecommand{\eprint}[2][]{\url{#2}}

\bibitem{ReichelMirrorMOT}
Reichel, J., H\"ansel, W. \& H\"ansch, T. W. 
\newblock \bibinfo{journal}{\bibinfo{title}{Atomic micromanipulation
  with magnetic surface traps}}.
\newblock {\emph{\JournalTitle{Phys. Rev. Lett.}}}
  \textbf{\bibinfo{volume}{83}}, \bibinfo{pages}{3398--3401}
  (\bibinfo{year}{1999}).


\bibitem{Fol00}
\bibinfo{author}{Folman, R.} \emph{et~al.}
\newblock \bibinfo{journal}{\bibinfo{title}{Controlling cold atoms using
  nanofabricated surfaces: Atom chips}}.
\newblock {\emph{\JournalTitle{Phys. Rev. Lett.}}}
  \textbf{\bibinfo{volume}{84}}, \bibinfo{pages}{4749--4752}
  (\bibinfo{year}{2000}).
% \newblock \doiprefix 10.1103/PhysRevLett.84.4749.

\bibitem{Folman:ChipReview}
\bibinfo{author}{Folman, R.}, \bibinfo{author}{Krüger, P.},
  \bibinfo{author}{Schmiedmayer, J.}, \bibinfo{author}{Denschlag, J.} \&
  \bibinfo{author}{Henkel, C.}
\newblock \bibinfo{journal}{\bibinfo{title}{Microscopic atom optics: From wires
  to an atom chip}}.
\newblock {\emph{\JournalTitle{Advances In Atomic, Molecular, and Optical
  Physics}}} \textbf{\bibinfo{volume}{48}}, \bibinfo{pages}{263 -- 356}
  (\bibinfo{year}{2002}).
\newblock
  % \urlprefix\url{http://www.sciencedirect.com/science/article/pii/S1049250X02800118}.
% \newblock \doiprefix http://dx.doi.org/10.1016/S1049-250X(02)80011-8.

\bibitem{reichel2011}
Reichel, J. \& Vuletic, V. {\em Atom chips\/}. (Wiley-VCH, 2011).

\bibitem{Reichel:2002}
\bibinfo{author}{Reichel, J.}
\newblock \bibinfo{journal}{\bibinfo{title}{Microchip traps and Bose-Einstein
  condensation}}.
\newblock {\emph{\JournalTitle{Applied Physics B}}}
  \textbf{\bibinfo{volume}{74}}, \bibinfo{pages}{469--487}
  (\bibinfo{year}{2002}).
% \newblock \doiprefix 10.1007/s003400200861.

\bibitem{Fortagh2007}
\bibinfo{author}{Fort\'agh, J.} \& \bibinfo{author}{Zimmermann, C.}
\newblock \bibinfo{journal}{\bibinfo{title}{Magnetic microtraps for ultracold
  atoms}}.
\newblock {\emph{\JournalTitle{Rev. Mod. Phys.}}}
  \textbf{\bibinfo{volume}{79}}, \bibinfo{pages}{235--289}
  (\bibinfo{year}{2007}).
% \newblock \doiprefix 10.1103/RevModPhys.79.235.

\bibitem{Kru03}
\bibinfo{author}{Kr\"uger, P.} \emph{et~al.}
\newblock \bibinfo{journal}{\bibinfo{title}{Trapping and manipulating neutral
  atoms with electrostatic fields}}.
\newblock {\emph{\JournalTitle{Phys. Rev. Lett.}}}
  \textbf{\bibinfo{volume}{91}}, \bibinfo{pages}{233201}
  (\bibinfo{year}{2003}).
% \newblock \doiprefix 10.1103/PhysRevLett.91.233201.

\bibitem{Han01}
\bibinfo{author}{H\"ansel, W.}, \bibinfo{author}{Reichel, J.},
  \bibinfo{author}{Hommelhoff, P.} \& \bibinfo{author}{H\"ansch, T.~W.}
\newblock \bibinfo{journal}{\bibinfo{title}{Magnetic conveyor belt for
  transporting and merging trapped atom clouds}}.
\newblock {\emph{\JournalTitle{Phys. Rev. Lett.}}}
  \textbf{\bibinfo{volume}{86}}, \bibinfo{pages}{608--611}
  (\bibinfo{year}{2001}).
% \newblock \doiprefix 10.1103/PhysRevLett.86.608.

\bibitem{Cas00}
\bibinfo{author}{Cassettari, D.}, \bibinfo{author}{Hessmo, B.},
  \bibinfo{author}{Folman, R.}, \bibinfo{author}{Maier, T.} \&
  \bibinfo{author}{Schmiedmayer, J.}
\newblock \bibinfo{journal}{\bibinfo{title}{Beam splitter for guided atoms}}.
\newblock {\emph{\JournalTitle{Phys. Rev. Lett.}}}
  \textbf{\bibinfo{volume}{85}}, \bibinfo{pages}{5483--5487}
  (\bibinfo{year}{2000}).
% \newblock \doiprefix 10.1103/PhysRevLett.85.5483.

\bibitem{Ott01}
\bibinfo{author}{Ott, H.}, \bibinfo{author}{Fortagh, J.},
  \bibinfo{author}{Schlotterbeck, G.}, \bibinfo{author}{Grossmann, A.} \&
  \bibinfo{author}{Zimmermann, C.}
\newblock \bibinfo{journal}{\bibinfo{title}{Bose-Einstein condensation in a
  surface microtrap}}.
\newblock {\emph{\JournalTitle{Phys. Rev. Lett.}}}
  \textbf{\bibinfo{volume}{87}}, \bibinfo{pages}{230401}
  (\bibinfo{year}{2001}).
% \newblock \doiprefix 10.1103/PhysRevLett.87.230401.

\bibitem{Hae01}
\bibinfo{author}{H\"ansel, W.}, \bibinfo{author}{Hommelhoff, P.},
  \bibinfo{author}{H\"ansch, T.} \& \bibinfo{author}{Reichel, J.}
\newblock \bibinfo{journal}{\bibinfo{title}{{B}ose-{E}instein condensation on a
  microelectronic chip}}.
\newblock {\emph{\JournalTitle{Nature}}} \textbf{\bibinfo{volume}{413}},
  \bibinfo{pages}{498} (\bibinfo{year}{2001}).
% \newblock \doiprefix 10.1038/35097032.

\bibitem{Sch03}
\bibinfo{author}{Schneider, S.} \emph{et~al.}
\newblock \bibinfo{journal}{\bibinfo{title}{Bose-Einstein condensation in a
  simple microtrap}}.
\newblock {\emph{\JournalTitle{Phys. Rev. A}}} \textbf{\bibinfo{volume}{67}},
  \bibinfo{pages}{023612} (\bibinfo{year}{2003}).
% \newblock \doiprefix 10.1103/PhysRevA.67.023612.

\bibitem{Sch05}
\bibinfo{author}{Schumm, T.} \emph{et~al.}
\newblock \bibinfo{journal}{\bibinfo{title}{{M}atter-wave interferometry in a
  double well on an atom chip}}.
\newblock {\emph{\JournalTitle{Nature Physics}}} \textbf{\bibinfo{volume}{1}},
  \bibinfo{pages}{57--62} (\bibinfo{year}{2005}).
% \newblock \doiprefix 10.1038/nphys125.

\bibitem{Bohi:2009ku}
\bibinfo{author}{B{\"o}hi, P.} \emph{et~al.}
\newblock \bibinfo{journal}{\bibinfo{title}{{Coherent manipulation of
  Bose-Einstein condensates with state-dependent microwave potentials on an
  atom chip}}}.
\newblock {\emph{\JournalTitle{Nature Physics}}} \textbf{\bibinfo{volume}{5}},
  \bibinfo{pages}{592--597} (\bibinfo{year}{2009}).
% \newblock \doiprefix 10.1038/nphys1329.

\bibitem{Gallego:2009ww}
\bibinfo{author}{Gallego, D.}, \bibinfo{author}{Hofferberth, S.},
  \bibinfo{author}{Schumm, T.}, \bibinfo{author}{Kr\"{u}ger, P.} \&
  \bibinfo{author}{Schmiedmayer, J.}
\newblock \bibinfo{journal}{\bibinfo{title}{Optical lattice on an atom chip}}.
\newblock {\emph{\JournalTitle{Opt. Lett.}}} \textbf{\bibinfo{volume}{34}},
  \bibinfo{pages}{3463--3465} (\bibinfo{year}{2009}).
% \newblock \doiprefix 10.1364/OL.34.003463.

\bibitem{leung2011}
Leung, V~Y., Tauschinsky, A., Van~Druten, N. \& Spreeuw, R~J. 
\newblock \bibinfo{journal}{\bibinfo{title}{Microtrap arrays on magnetic film atom chips for quantum information science}}.
\newblock {\emph{\JournalTitle{Quantum Information Processing}}} \textbf{\bibinfo{volume}{10}},
  \bibinfo{pages}{955} (\bibinfo{year}{2011}).



\bibitem{jose2014}
Jose, S., Surendran, P., Wang, Y., Herrera, I., Krzemien, L., Whitlock, S., McLean, R., Sidorov, A. \& Hannaford, P.
\newblock \bibinfo{journal}{\bibinfo{title}{Periodic array of Bose-Einstein condensates in a magnetic lattice}}.
\newblock {\emph{\JournalTitle{Physical Review A}}} \textbf{\bibinfo{volume}{89}},
  \bibinfo{pages}{051602} (\bibinfo{year}{2009}).
  

\bibitem{DiVincenzo:2000dn}
\bibinfo{author}{DiVincenzo, D.~P.}
\newblock \bibinfo{journal}{\bibinfo{title}{The physical implementation of
  quantum computation}}.
\newblock {\emph{\JournalTitle{Fortschritte der Physik}}}
  \textbf{\bibinfo{volume}{48}}, \bibinfo{pages}{771--783}
  (\bibinfo{year}{2000}).

\bibitem{Gunther2005}
\bibinfo{author}{G\"unther, A.} \emph{et~al.}
\newblock \bibinfo{journal}{\bibinfo{title}{Combined chips for atom optics}}.
\newblock {\emph{\JournalTitle{Phys. Rev. A}}} \textbf{\bibinfo{volume}{71}},
  \bibinfo{pages}{063619} (\bibinfo{year}{2005}).
% \newblock \doiprefix 10.1103/PhysRevA.71.063619.

\bibitem{Gierling:2011ta}
\bibinfo{author}{Gierling, M.} \emph{et~al.}
\newblock \bibinfo{journal}{\bibinfo{title}{Cold-atom scanning probe
  microscopy}}.
\newblock {\emph{\JournalTitle{Nature nanotechnology}}}
  \textbf{\bibinfo{volume}{6}}, \bibinfo{pages}{446--451}
  (\bibinfo{year}{2011}).
% \newblock \doiprefix 10.1038/nnano.2011.80.

\bibitem{Roy16}
\bibinfo{author}{Roy, R.}
\newblock \emph{\bibinfo{title}{An Integrated Atom Chip for The Detection and
  Manipulation of Cold Atoms Using a Two-photon Transition}}.
\newblock Ph.D. thesis, \bibinfo{school}{Centre for Quantum Technologies,
  National University of Singapore} (\bibinfo{year}{2015}).

\end{thebibliography}
\end{document}